\documentstyle[psfig]{lamuphys}
\makeatletter
\let\chapter\hid@chapter
\makeatother
\begin{document}
\pagenumbering{arabic}
\title{Quantitative morphology and color gradients of E+A galaxies in distant
galaxy clusters}
\titlerunning {Color gradients of E+A galaxies in distant galaxy clusters}

\author{Paola Belloni}

\institute{Universit\"atssternwarte, Scheinerstra\ss e 1,
D--81679 M\"unchen, Germany}
\maketitle

Three basic scenarios have been invocated to explain
the sudden rise and decline of star formation in distant clusters
that lead to the transformation of ``active'' galaxies
 into the elliptical$-$ and  S0$-$types which today dominate rich clusters,
 i.e. the so called Butcher-Oemler effect, (Butcher \& Oemler,1978).
These are: galaxy interactions or mergers with nearly equal mass neighbors
(Lavery \& Henry, 1988),
 gas-rich field galaxies running for the first time
into the hot intercluster gas  which ignites a brief but energetic
 episode of star formation
(Bothun \& Dressler, 1986), and
high speed close encounters of gas-rich galaxies resulting in
 non-disruptive interactions or ``galaxy harassment'',
(Moore, 1996).

Although neither of the  proposed mechanisms for triggering
 star formation has a definitive answer in its favour,
the high resolution of the HST images of distant clusters has recently
provided  a unique tool to face this question.
Indeed, it allows us to study the  spatial distribution of the
 starburst and thus to distinguish among the physical mechanisms
 at its origin.
In the case  of an infalling or harassed spiral the enhanced
star formation would likely be a  galaxy-wide phenomenon,
or confined to the disk. If, on the other hand, the original
 galaxy was an elliptical that accreted gas from a dwarf galaxy,
 the burst signatures should be detectable as a bluer color of
the nuclear region. 
Finally, in interacting or merging disk galaxies both behaviours
have been observed: galaxies with starburst concentrated to
 the very center (Scoville et al.,1991) and others that
 show enhanced star formation  on a galaxy-wide scale (Standford, 1991).

We focuse on one class   of ``active'' galaxies,
 the post-starburst or E+A galaxies, the latter being
 {\it only} a description of the spectra
which appear to have an A-star component added to an old elliptical
like component. These spectral features  are interpreted 
 as evidence of a recent ($<$ 1.5 Gyr) burst of star formation.
Ground based narrow-band  photometry of 4 intermediate redshift
clusters at z=0.4-0.5 has provided us with a sample of 73 E+A galaxies
being secure cluster members (Belloni et al.,1995; Belloni \& R\"oser, 1996).
This is the largest sample of such galaxies
 found to date in  distant clusters.
Cluster membership and spectral type have been obtained by fitting 
the observed low-resolution spectral energy distributions
with template spectra built up with Bruzual \& Charlot (1997) population
synthesis models. They represent the temporal evolution 
of a strong star formation episode (involving 20\% of the
original galactic mass)  in an elliptical galaxy or
in a spiral galaxy with star formation truncated after the burst. 
Our approach allows us to detect bursts younger than 2 Gyr
while galaxies with an older burst will not be
distinguished from a passively evolving elliptical galaxy.

Due to the small HST field of view only for 33  E+A galaxies
 HST images are available (WFPC2). We have retrieved them and determined:
 a) {\bf surface brightness  profiles} for all
 E+A galaxies but those showing highly irregular morphology. 
  A fit with an exponential or a r$^{-1/4}$ law has been performed and 
for disk galaxies a Hubble type classification has been attempted
 on the basis of the buldge to disk ratio,
b) {\bf color gradients} for the 10  E+A galaxies in
 Cl0016+16 (observed in the F555W and F814W filters). 
{\bf
\begin{table}
\caption[]{Results of the surface brightness analysis of 33 E+A galaxies
in intermediate redshift clusters. The E+A fraction is given with
respect to the secure cluster members,
about 120 galaxies brighter than m$_{R}$=22.5 per cluster.
For merging/interacting galaxies we indicate if
 possible  whether they appear disk or bulge dominated.}
\begin{center}
\begin{tabular}{lccccccc}
Cluster & \,\, & \,\,E+A\,\, & \,\,E+A\,\,  & \,\, Morph.\,\, & \,\,Morph.\,\, & \,\,Morph.\,\,
 & \,\, Merger\,\, \\ [0.6ex]
  &  &  & (HST)   & Elliptical  &  S0-Sa & Sb-Sc &  Interact.   \\
\hline \\  
Cl0016+16  &  & 20 (21\%) & 11  &  1 & 3 & 6 & 1    \\
(z=0.54)  &  & &  &  &  & & Disk-dominated  \\
& & & & & & &  \\
Cl0939+47 & & 35 (22\%) & 11 & & & 7 & 4  \\
(z=0.41)  & & & & & & &   \\
& & & & & &  & \\
Cl0303+17 & & 22 (22\%) & 7 & & 1 & 5 & 1  \\
(z=0.41) & & & & & & & Irregular  \\
& & & & & &  & \\
Cl1447+26 & & 8 (9\%) &  4 & & 1 & 1 &  2 \\
(z=0.38) & & & & & & &  Disk-dominated \\
\end{tabular}
\end{center}
\end{table}
}

We found that {\bf most of the galaxies and all those
showing signs of interaction are disk systems} judging from the
 exponential nature of the profiles (Tab.1). 
In our sample only one  E+A galaxy  has a regular elliptical profile.
We also found that the E+A galaxies in Cl0016+16  show little spatial variation
 in V-I colors (restframe U-V) (Fig.1).
Only one young E+A galaxy with estimated post-starburst age  0.8 Gyr
has a starburst predominantly located in the nucleus 
whereas a second one has a weaker evidence of the same behaviour.
However, the E+A color gradients
are flatter compared to those of cluster ellipticals
(Belloni et al, 1997) that are consistent with the
 $\Delta$(U-R)/$\Delta$log r=$-$0.23 mag
  observed in the nearby ones (Franx et al.,1993).
Thus, our  large and homogeneous sample of
 distant E+A galaxies confirms the claim that their  {\bf  enhanced
 star formation is a galaxy-wide phenomenon.}
Indeed, similar results have been obtained
for the post-starburst galaxies in the Coma cluster by Caldwell et
 al. (1996).
We speculate that strong interactions did not create
 most of  the post-starburst galaxies  in
our representative sample of galaxy clusters at z=0.4$-$0.5.
Non-disruptive interactions among galaxies
or  interactions between galaxies in the cluster or subcluster
structures and the ICM are likely to be the dominant mechanism.

\begin{figure}[h]
\begin{center} \parbox{11.5cm}{ 
\psfig{figure=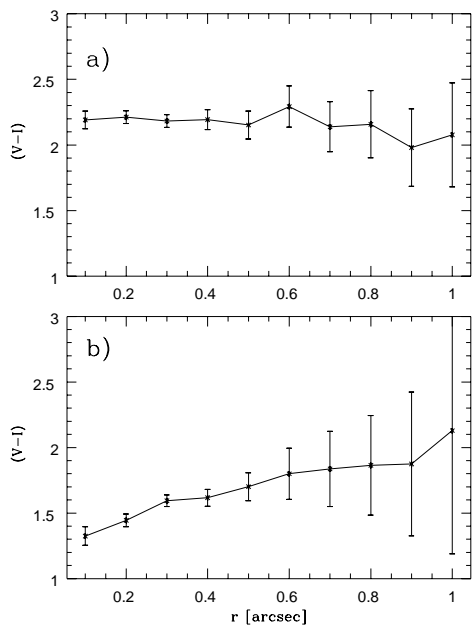,width=8.8cm}}
\caption[a]{\label{mgbscv} 
\,a) Typical color gradients of  E+A  galaxies
 in CL0016+16 (z=0.54).
 \,b) One of the two E+A galaxies in Cl0016+16 showing strong color gradients.
The very blue central colors (typical of spirals) indicate that the
 starburst is located in the nucleus. 
1$''$= 7.3 Kpc (H$_o$=50 km sec$^{-1}$ Mpc$^{-1}$ and q$_0$=0)}
\end{center} 
\end{figure}

%
%
%

\end{document}